\documentstyle[preprint,aps]{revtex}
\begin{document}
\tightenlines
\title{Asymmetric emission of high energy electrons in the two-dimensional hydrodynamic expansion of large xenon clusters irradiated by intense laser fields}   
\author{V. Kumarappan, M. Krishnamurthy, and D. Mathur}
\address{Tata Institute of Fundamental Research, 1 Homi Bhabha 
Road, Mumbai 400 005, India.}
\date{\today}
\maketitle
\begin{abstract}
Energy spectra and angular distributions have been measured of electrons that are emitted upon disassembly of $Xe_{150000}$ following irradiation by intense (10$^{15}$-10$^{16}$ W cm$^{-2}$) laser pulses whose durations are varied over the 100-2200 fs range. The cluster explosion dynamics occur in the hydrodynamic regime. Electron emission is found to be unexpectedly asymmetric and exhibits a resonance when the laser pulse duration is $\sim$1 ps. These results are rationalized by extending the hydrodynamic model to also take into account the force that the light field exerts on the polarization charge that is induced on surface of the cluster. We show that the magnitude of this electrostrictive force is comparable to those of Coulombic and the hydrodynamic forces, and it exhibits resonance behavior. Contrary to earlier understanding, we find that low-energy electrons are connected to the resonance in energy absorption by the cluster. The high-energy electrons seem to be produced by a mechanism that is not so strongly influenced by the resonance. 
\end{abstract}
\pacs{52.50.Jm, 34.80 Kw, 36.40.c, 52.25.Nr }

\section{Introduction}

Ready availability of intense, ultrashort laser sources has driven widespread interest in the response of various forms of matter to intense light. The focus of the bulk of contemporary research has been either on dynamics involving low-density matter, in the form of isolated atoms or molecules, or on interactions that involve high-density matter in the form of solids. Investigations have lead to the discovery of several fundamental phenomena in both types of matter, and many avenues for technological applications have opened up. Gas-phase clusters are uniquely placed between the two density regimes. The large local density in a cluster gives rise to solid-like behavior but, at the same time, sizes of clusters are small enough to ensure that atomic and molecular effects do not lose their relevance. This co-mingling of two separate aspects of laser-matter interaction is one of the main reasons for growing interest in the fundamental aspects of the physics of cluster dynamics in intense laser fields. Another, very important facet of laser-cluster interactions that has fueled research in the field is the discovery of the extraordinarily high efficiency with which large rare-gas clusters absorb laser energy \cite{absorption} and subsequently redistribute it into high energy ions \cite{natureions}, electrons \cite{electronprl,nam}, and photons \cite{naturexray}. The efficient conversion of laser energy into high energy particles and radiation has many important implications from the viewpoint of applications. The more notable among these are coherent radiation in the x-ray regime through high-harmonic generation \cite{hhg}, intense incoherent x-rays emitted by hot plasmas \cite{xrays}, high-energy electrons by laser wakefield acceleration \cite{lwfa}, and, recently, generation of ultrashort pulses of monoenergetic neutrons from nuclear fusion in cluster plasmas \cite{fusion}. Moreover, there are a plethora of other potential applications in extreme ultraviolet (EUV) lithography, time-resolved neutron diffraction, and table-top accelerators that await exploitation. However, despite the fundamental importance of laser-cluster interactions, and many tantalizing applications, proper insight into the dynamics of energy absorption and redistribution remains elusive. The complexity of the interaction necessitates the use of various approximations in modeling observations. Experimental and theoretical investigations continue to remain necessary in order to verify the validity of different schemes.  

There exist two major theoretical models that seek to explain the evolution of the cluster under intense field irradiation. Neither of them, the hydrodynamic expansion model or the ionization ignition - Coulomb explosion model, can adequately explain the experimentally observed features that presently drive research in this area. In both models, atoms in a cluster are tunnel ionized at the leading edge of the incident laser pulse. As the ionized electrons are removed from the cluster, what gets left behind is a positively charged core that gives rise to an increasing potential barrier to further removal of electrons. The question of whether the barrier is sufficient to retain a large fraction of electrons or not is the bone of contention between the two models. In the hydrodynamic expansion model, it is assumed that retention of most of the electrons by the cluster results in a spherically symmetric plasma. The retained electrons absorb energy from the laser by collisional inverse bremsstrahlung. The hot electron plasma expands due to hydrodynamic pressure, and transfers energy to the ions. The expansion velocity of the plasma is determined by the plasma sound speed, and since the ion and electron charge clouds expand at the same speed, the ions are expected to be significantly hotter than the electrons. On the other hand, the Coulomb explosion-ionization ignition model of the dynamics avers that electrons leave the cluster rapidly after tunnel ionization. As a result, there is a build-up of charge on the cluster that gives rise to a radial field that can be large enough to drive further ionization at the surface of the sphere. The removal of these electrons increases the radial field further and ``ignites" ionization. The cluster then explodes due to the Coulombic repulsion between the positively charged ions. At the most basic level, prevailing wisdom indicates that the hydrodynamic expansion model is expected to hold for clusters that are large enough in size to retain a substantial fraction of the electrons, while the ionization ignition-Coulomb explosion model requires the prompt removal of most of the electrons, which would be the case for small clusters. In this paper we focus attention on the hydrodynamic expansion interpretation of cluster laser dynamics as our experiments are concerned with large clusters of xenon, $Xe_n$ ($n\sim$150000).

Although the hydrodynamic expansion model \cite{plasma} remains the most successful effort at quantitative description of the interaction of large clusters with intense laser fields, it is, nevertheless, recognized that the model has shortcomings; some attempts have recently been made to improve upon it \cite{liu,milchberg}. Due to the complexity of the problem, the original model, as well as various modifications, have hitherto assumed that the plasma evolution occurs in spherically symmetry fashion. However, spherical symmetry has raised problems when the model has been sought to be used in order to provide adequate quantitative descriptions of the dynamics that give rise to the pronounced asymmetry that is  experimentally observed in the energy spectra of emitted ions following cluster disassembly \cite{argon,xenon}. Results of these experiments have shown that {\em asymmetric} emission of the fastest (most energetic) ions is closely linked to the efficiency of energy absorption in the cluster, and that a resonance in the latter leads to enhancement of the former. By measuring ion energies both along and perpendicular to the laser polarization vector as a function of pulse duration, it has been shown that ion energies show a resonance at a particular value of laser pulse duration cluster and, somewhat unexpectedly, that the  disassembly of the $Xe_{150000}$ cluster is most asymmetric when the mean ion energy is maximum. In order to account for this unexpected result, an extension has been proposed of the hydrodynamic model \cite{xenon} that takes cognizance of the force that the light field exerts on the induced polarization charge on the surface of the cluster. It has been shown that the magnitude of this force can be comparable to the Coulombic and hydrodynamic forces, and that it also shows resonance-like behavior. It can, thus, cause an asymmetry in the ion emission with the observed characteristics. 

The polarization force that has been proposed to rationalize the asymmetry of the high energy ions should also influence the energy spectrum, and the angular distribution, of the electrons that are ejected from the cluster upon its disassembly. We present in the following experimental data to show that this is, indeed, the case. There has been only one previously reported study of electron emission from large clusters that explode in the hydrodynamic expansion regime \cite{electronprl} that, in discussing low-and high-energy components in the electron energy distribution, have a direct bearing on the questions of resonance and asymmetry that we address in this paper. Surprisingly, contrary to what might be expected on the basis of this earlier measurement of the electron energy spectrum \cite{electronprl}, we find that low-energy electrons are connected to the resonance in the energy absorption. The high-energy electrons seem to be produced by a mechanism that is not influenced by the resonance. In the course of this work, another report has appeared \cite{nam} in which results are presented that indicate that the high energy component might extend to values as high as 500 keV when argon and xenon clusters are irradiated by laser intensities that are in the 10$^{17}$ W cm$^{-2}$ range. We discuss our explanation of the electron data in the context of recently developed understanding of the role of resonance in determining electron and ion energy distributions. Both the experimental data and the proposed extension of the hydrodynamic model emphasize the need for a two-dimensional description of the dynamics of cluster-laser interaction that treats the charge and the field distributions in self-consistent fashion.    

\section{Experimental method}
  
Our experimental setup is similar to the one described recently \cite{argon,argon_xrays}, with some modifications. It is shown schematically in Fig. 1. $Xe_{150000}$ clusters were produced using a solenoid-driven pulsed nozzle with a 500 ${\mu}$m throat diameter, and the centerline beam was transfered to a high vacuum chamber via a $55^{\circ}$ nickel skimmer with a 250 ${\mu}$m diameter aperture. Xenon cluster sizes were estimated using Hagena's scaling law \cite{hagena}. A dual microchannel plate (MCP) assembly was used at the end of a 19 cm time-of-flight (TOF) spectrometer to detect the ions and the electrons. Ion energies were determined by converting the field-free ion time-of-flight to energy. Electron energies were measured by scanning the voltage on a retarding potential analyzer from zero to -5 kV, integrating the signal near $t=0$ over time, and then differentiating the yield thus obtained with respect to the voltage. The MCP signal was recorded using a 500 MHz, 1 GS s$^{-1}$ digital oscilloscope.  

The Ti:sapphire laser used can generate 55 mJ per pulse, with 100 fs pulsewidth, but only upto 12 mJ was used for this series of experiments. With 100 fs pulses, typical intensities that were accessed in this series of experiments covered the range 10$^{14}$ - 10$^{16}$ W cm$^{-2}$. The pulse duration of the laser was varied by changing the grating separation in the compressor. The actual pulse duration was measured using an second-harmonic generation autocorrelator located just outside the vacuum chamber. At the intensities used here, we did not find appreciable effects of chirp on the results that are presented below. The laser pulse energy was kept constant for these measurements; hence, the intensity falls linearly as the pulse duration is increased. While this procedure does imply that two variables - the pulse duration and the laser intensity - are simultaneously changed, it is known that the dependence of energy absorption on intensity is not very strong \cite{springate}. Moreover, earlier experimental work \cite{xenon} has shown that the existence of a maximum in the ion energy as the laser intensity is reduced. This strengthens the case for the importance of the pulse duration, and, by implication, of the resonance that has been observed in ion energy spectra.

\section{Role of resonance in the hydrodynamic model}

Before we present the results of our experiments on electron emission upon disassembly of large xenon clusters, it is pertinent to briefly review the role of resonance in attempts that seek to rationalize the observed asymmetries in ion emission in the case of $Xe_{150000}$ clusters at laser intensities in the 10$^{16}$ W cm$^{-2}$ range \cite{xenon}. Our electron data can then be properly placed in the context of the most recent developments of the hydrodynamic expansion model. 

Laser energy is deposited into the cluster through inverse bremsstrahlung. The hydrodynamic model assumes that the cluster can be described as a uniform and isotropic plasma with a single electron temperature. If we take the plasma to be a dielectric sphere that is placed in a uniform, external electric field, it is known that the application of the field polarizes the medium, and alters the field in and around the sphere (see Fig. 2). Under the assumption of spherical symmetry, together with the Drude approximation for the dielectric constant of the plasma, the electric field distribution inside and outside the sphere can be found using the Laplace equation with appropriate boundary conditions. Furthermore, if we assume that the plasma dimensions are much smaller than the wavelength of the laser light (780 nm, in our case, compared to Xe-cluster sizes that are on the order of 10 nm), the electric field can be taken to be uniform over the entire cluster and the problem can be analytically solved. The solution is available in several textbooks on electromagnetic theory \cite{jackson}. For an applied electric field, $E_oe^{i\omega t}$, and a dielectric constant, $\epsilon$, the electric field within the dielectric sphere, $E_{in}(r)$, and outside it, $E_{out}(r)$, can be expressed, respectively, as
\begin{equation}
E_{in}(r) = \frac{3}{\epsilon + 2} E_o e^{i\omega t} cos(\theta),
\end{equation}
\begin{equation}
E_{out}(r) = \left(\frac{2a^3}{r^3}\frac{\epsilon - 1}{\epsilon + 2} + 1 \right) E_o e^{i\omega t} cos(\theta).
\end{equation}
These fields are the sum of the applied field and the field due to an induced dipole on the dielectric sphere. It is known that for all homogeneous ellipsoids, the field inside is uniform, and this is reflected in the expression for $E_{in}(r)$. An induced surface polarization charge produces the dipole field; the corresponding charge density, $\sigma_{pol}$, is
\begin{equation}
\sigma_{pol} = \frac3{4\pi}\left(\frac{\epsilon - 1}{\epsilon + 2}\right) E_o e^{i\omega t} cos(\theta).
\end{equation}
We note that $\epsilon$ in these equation represents a complex quantity, and hence the fields and the surface charge density are not in phase with the applied laser field.  The description of the cluster plasma in terms of these is valid only for a rigid dielectric sphere - the deformation of the sphere due to electrostrictive forces has been not been considered. The pressure on the surface can be estimated  by calculating the force that is exerted on a surface element by multiplying the charge density by the electric field at that point. We took care to exclude the field due to the charge element itself by taking the average of the fields on the two sides of the surface element as the effective field. The charge on the surface element makes equal but opposite contributions to the fields on the two sides, and the averaging removes this contribution from the field. The cycle-averaged force calculated in this manner is
\begin{equation}
P_{pol} = \frac{1}{2} Re (\vec{\sigma}_{pol}\cdot\vec{E}^{\ast}_{ave}) = \frac{9}{16\pi^2}\frac{|\epsilon|^2 - 1}{|\epsilon + 2|^2} E_o^2 cos^2\theta.
\end{equation}
Using the expression for the electric field within the cluster, that the rate of energy deposition by an applied field is expressed as
\begin{equation}
\frac{\delta U}{\delta t} = \frac{9\omega}{8\pi}\frac{Im(\epsilon)}{|\epsilon + 2|^2}|E_o|^2.
\end{equation}
The absorption of laser energy by the cluster goes through a resonance at Re$(\epsilon)$ = -2, where the field inside the cluster reaches a maximum value (see Fig. 3). The figure also shows that the polarization-dependent pressure also exhibits similar, resonance-like behavior, although at a slightly shifted value of $\epsilon$. Large negative values of Re$(\epsilon)$ correspond to large electron densities, which are present in the early part of the laser pulse. As time evolves and the cluster expands, the electron density falls and the real part of the dielectric constant, which initially has large negative values (Fig. 3), approaches -2 where both the electric field within the sphere and the rate of energy absorption, are resonantly enhanced. The functional dependences shown in Fig. 3 are for a laser intensity of 1$\times$10$^{15}$ W cm$^{-2}$; the imaginary part of $\epsilon$ was chosen such that the electric field was enhanced 5-fold at resonance.
 
It is of interest to compare the magnitude of the electrostrictive pressure due to the applied laser field with typical Coulombic and hydrodynamic pressures. In order to extract some quantitative insight, we used the same plasma parameters as were earlier considered by Ditmire {\it et al.} \cite{plasma}: the electron density was taken to be 5.4 $\times$ 10$^{21}$ cm$^3$ (which is three times the critical density for 800 nm light), the temperature was assumed to be 1 keV, the radius of the cluster plasma was 100 {\AA}, and the total charge on the cluster was 105. Under these conditions, both the Coulombic pressure and the hydrodynamic pressure have a magnitude of $\sim$3$\times$10$^{11}$ N m$^{-2}$. It is therefore clear that the polarization pressure that we discuss above, values of which are on the same order of magnitude (see Fig. 3), does, indeed, contribute to the expansion dynamics of the cluster plasma. As such, the contribution of the polarization pressure must be included in any realistic modeling of the evolution of the system. Consideration of this pressure has been shown to qualitative explain the asymmetric ion emission that occurs upon disassembly of $Xe_{150000}$ clusters upon irradiation by laser light of intensity 10$^{16}$ W cm$^{-2}$ \cite{xenon}. 

We note that the polarization pressure survives cycle averaging, and is directed outwards from the sphere (the pressure is positive) over a large range of values for the real part of $\epsilon$. As shown in Fig. 3, even in the region where this pressure does become negative, its magnitude remains small in comparison with the positive pressure near the resonance. The pressure also acts on the entire cluster plasma, causing both the ionic and the electronic components to expand asymmetrically. This actually follows from a built-in assumption in the model: the ions and the electrons are uniformly distributed and occupy the same volume at all points of time. The only exception is that the electrons are allowed to free-stream out of the cluster if they have sufficient energy. But once free, these electrons are no longer part of the dielectric plasma described by the equations that we have considered so far.

\section{Results and discussion}

Figure 4 shows a typical, raw MCP signal corresponding to the electrons emitted by the cluster in our experiments. The waveform was recorded using a 500 MHz, 1 GS s$^{-1}$ digital storage oscilloscope. The front end of the MCP was kept at 500 V for this measurement. The zero on the time scale was determined by measuring the arrival time of the laser pulse with a fast photodiode. Three distinct features, marked A, B and C, are clearly seen in the signal. The dip in the MCP output that is marked A is the fastest signal that can readily be distinguished from noise. It occurs within a couple of nanoseconds after the arrival of the laser pulse. This signal could be either due to very fast electrons, possessing energies in excess of 10 keV, or to photons that are either emitted or scattered by the clusters. As discussed in the following, it was not possible for us to distinguish between these two possibilities in totally unambiguous fashion in the present experimental setup. Feature B, on the other hand, is known to be the signal from electrons that possess energies up to 5 keV. We shall refer to this as the peak that is due to ``warm electrons". All measurements of electron yield and asymmetry that are reported in the following pertain to the integrated signal of peaks A and B. Feature C in Fig. 4 is due to imperfect impedance matching between the detector and the oscilloscope that gives rise to ``ringing" in the electron signal.    

Electrons with a few keV kinetic energy are too fast for time-of-flight measurement. Here we make use of the retarding potential energy analyzer (RPA) that is placed before the detector at the end of our time-of-flight spectrometer in order to obtain electron energy spectra. As the (negative) voltage on the RPA is increased, the integrated electron signal is measured at each voltage. Since the RPA rejects all electrons with energy less the applied voltage, $V_{RP}$, the measured signal is given by 
\begin{equation}
S_{MCP}(V_{RP}) = \int_{V_{RP}}^\infty f(V) dV.
\end{equation}
The actual distribution of electron energies $f(V)$ can then be obtained by simply differentiating the integrated electron signal with respect to V. Figure 5 shows an electron energy spectrum obtained by this method for $Xe_{25000}$ clusters irradiated by laser light of intensity 8$\times$10$^{15}$ W cm$^{-2}$. The polarization vector of the incident light was directed along the axis of our time-of-flight spectrometer. We note that a single temperature of 700 eV adequately describes the electron energy profile in this case. The fast peak designated A in Fig. 4 could not be resolved in these experiments with as much as -5 kV applied on the RPA. If the A-peak is to be attributed to the ``hot electrons" peak that was detected in the spectrum reported by Shao {\it al.} \cite{electronprl}, then data shown in Fig. 5 indicate that the electron energy would have to be much larger than the 2-3 keV that they claimed. This might be due to the fact that the experiments conducted by Shao {\it et al.} were on much smaller $Xe_n$ clusters, comprising $\sim$2000 atoms. 

In the case of clusters of larger size, such as $Xe_{150000}$, the integrated electron yields that we measured as a function of RPA voltage are shown in Fig. 6 for two, mutually orthogonal, laser polarization directions. The actual spectrum can be obtained by differentiating such a data set, but the procedure is prone to amplifying noise, and hence raw, undifferentiated data is shown. Unlike in Fig. 5, in Fig. 6 it is possible to discern the distribution of electrons between the ``hot" and the ``warm" peaks. For instance, the integrated yield of electrons above 5 keV energy (with some fraction of this contribution coming from photons) is given by the data point at the far right of the diagram, which is approximately 3 units on the arbitrary scale. The integrated yield above 500 eV is represented by the data point on the far left of the diagram, and this is about 14 units. Thus the ``hot" electrons constitute 20\%, or less, of the total electron emission from the cluster along the laser polarization vector. By the same argument, the fraction for the perpendicular direction is about 50\%. On rotating the light polarization vector by 90$^\circ$ in our experiments, we found that the fast peak remains unchanged, while the warm electrons are considerably reduced in intensity. This in line with expectations of the modified hydrodynamic model that we discussed above, and lends support to our attribution of the peak to ``warm" electrons. We note the disagreement between our data and the only prior report on electron spectra \cite{electronprl} in which the warm electron yield seemed to go to zero when the laser polarization was in the perpendicular direction.

In order to probe the influence of the resonance and the electrostrictive force on the electron spectrum, we recorded the electron as the laser pulse duration was varied. We show in Fig. 7 typical MCP output pulses measured with different values of laser pulse duration that cover the range from 200 fs to 2.1 ps. We note that there is only marginal variation in the amplitude of peak A. On the other hand, the amplitude of peak B shows very substantial variations as the laser pulse duration is altered. This facet of our results is shown more clearly in the dependence we measure of the integrated electron yields on laser pulse duration, as shown in Fig. 8 for two, mutually orthogonal, laser polarization directions. A distinct resonance is observed when the laser pulse duration is $\sim$1.2 ps in the case when the laser polarization is parallel to the spectrometer axis. We measure a six-fold increase in the electron yield at resonance. The dependence for perpendicular polarization is much weaker. The asymmetry in the electron emission, measured as the difference between the yields obtained with parallel and perpendicular polarizations, also shows a resonance in much the same fashion that the corresponding asymmetry in ion emission did \cite{xenon}. 

As noted above, analysis of the MCP outputs (of the type shown in Fig. 7) at different values of pulse duration makes clear that the resonance in energy absorption is connected to the warm electrons, and not to the hot ones. In fact the hot electron yield seems to be relatively weakly dependent on the pulse duration. While this observation is in good agreement with the expectations from the electrostrictive force discussed above, it is in strong disagreement with the explanation offered by Shao {\it et al.} \cite{electronprl} for their electron data. They postulate that it is the extensive collisional heating of the plasma that produces the hot electrons during resonance, and that this heating process makes the angular distribution isotropic. On the other hand, the warm electrons are postulated to undergo a limited number of collisions. Hence, they retain the signature of above-threshold ionization (ATI) in their angular distribution which is known to produce a narrow distribution of electrons peaked along the laser polarization vector. Clearly, our experimental data rule out such an  interpretation. With the inclusion of the electrostrictive force, the observed variation with pulsewidth of the warm electrons is no longer unexpected. The resonance not only heats up the electrons, it also provides an additional, {\em asymmetric} lowering of the barrier that holds the electrons to the cluster. If the electron energy distribution is independent of position in the cluster, it is much more likely that a polar electron will free-stream out of the cluster rather than an equatorial one. Hence the asymmetry in the warm electron distribution is not a remnant of ATI, but a result of the asymmetry in the barrier that the electrons must overcome in order to leave the cluster! 

The interpretation of the ``warm" electron component that we offer on the basis of our data, and of the modified hydrodynamic expansion model that we have discussed, leaves unanswered the question of the ``hot" electrons. Since ``hot" electron production is no longer strongly associated with the resonance, new mechanisms must be found to explain both the energy and the angular distribution of these electrons. While the observed symmetry is probably due to the fact that these electrons have enough energy to free-stream directly from the interior of the cluster without requiring the binding potential to be lowered by the electrostrictive force, the source of these high-energy electrons remains to be discovered. We note here that it is somewhat surprising that a large fraction of electrons should remain unaffected by a large change in the pulse duration since the evolution of the cluster seems to be influenced very strongly by the position of the resonance. In particular, the ``fast" peak seems to remain unaltered as the pulse duration is increased well beyond the optimum value of 1.2 ps (in the case of $Xe_{150000}$), where not only is the peak intensity lower but the cluster has also expanded into a low density plasma before the peak light intensity is reached. This leads us to the very real possibility that the fast peak in our measurements of MCP signals consists of a not-insubstantial fraction of photons, which would automatically be isotropic. 

In the context of the two-dimensional hydrodynamic expansion, we note that one important difference between the forces acting upon the electrons and the ions in the cluster disassembly process lies in the directional properties of the different forces. While all the forces (Coulombic, hydrodynamic and electrostrictive) act outwards for positively-charged ions, the Coulomb force is directed inwards for the electrons. Consequently, in the case of electrons, the Coulombic force tends to counterbalance the effect of the other two forces. Due to the partial cancelation of the symmetric forces for the electrons, the influence of the asymmetric electrostrictive force should be seen more strongly in the electron than in the ion spectrum. This certainly seems to be borne out in the results of our experiments. For the same reason, the electron free-streaming rate should also be strongly modulated in favor of emission along the laser polarization vector. 

\section{Summary and conclusions}

We have studied the dynamics of $Xe_{150000}$ disassembly following irradiation by intense laser pulses whose duration is varied over the range from 200 fs to 2200 fs, with intensities in the range of 10$^{15}$ - 10$^{16}$ W cm$^{-2}$. The clusters are large enough for the explosion dynamics to occur in the hydrodynamic regime. Earlier work \cite{xenon} had established that the emission of high energy ions from such clusters is distinctly asymmetric. In order to account for this unexpected result, an extension of the hydrodynamic model was suggested that included the force that the light field exerts on the polarization charge that is induced on the surface of the cluster. It was shown that the magnitude of this force can be comparable to Coulombic and the hydrodynamic forces, and that the force also exhibits resonance-like behaviour. It can, thus, cause an asymmetry in the ion emission with the observed characteristics. Such a polarization force would also be expected to influence the energy spectrum and the angular distribution of the electrons ejected from the cluster, and we have presented experimental data here to show that this is, indeed, the case. 

In addition to providing data that adds veracity to the extension of the hydrodynamic model that has been proposed, our data also has some surprising facets. Contrary to earlier understanding of the electron energy spectrum, albeit established on the basis of one reported experiment \cite{electronprl}, we find that low-energy electrons are connected to the resonance in the energy absorption. The high-energy electrons seem to be produced by a mechanism that is not as strongly influenced by the resonance. We have placed our electron data in the context of recently proposed understanding of the role of resonance in determining electron and ion energy distributions. Both the experimental data and the proposed extension of the hydrodynamic model emphasize the need for a two-dimensional description of the dynamics of cluster-laser interaction that treats the charge and the field distributions self-consistently. 

The results presented here underscore the need for more sophisticated treatments that do not rely on spherical symmetry and uniform density of the plasma. Both these assumptions have been used frequently, largely because of the difficulty involved in doing away with them. Not only does our data directly question the appropriateness of the first assumption, the mechanism we have proposed here to explain our data is believed to be a first step towards a treatment that can deal with non-uniform charge distribution. A two-dimensional treatment, with self-consistent field and charge distribution, seems to be necessary for a satisfactory description of the dynamics. Milchberg {\it et al.}, who propose a self-consistent one-dimensional radial hydrodynamic model have made an attempt in this direction \cite{milchberg}. 

The possibility of simultaneous maximization of ion and electron energies, and asymmetry in the emission, may have profound implications. For instance, if the emission of deuterium ions can be made significantly anisotropic, the yield of fusion neutrons should become significantly higher due to a larger probability of energetic collisions \cite{fusion}. The directionality would also be useful in schemes for using cluster plasmas as sources of energetic highly charged ions. Even in applications where the primary interest is in the photons, for instance in EUV lithography, the ability to direct electrons and ions away from critical components should allow a reduction in damage due to energetic particle impact.

\begin{figure}
\caption{Schematic representation of the experimental apparatus for the measurement of electron spectra. HV: High voltage supply for the retarding potential analyzer, DAC: Digital-to-Analog-Convertor, DSO: Digital Storage Oscilloscope, GPIB: General Purpose Interface Board.}
\end{figure}

\begin{figure}
\caption{In the standard hydrodynamic expansion model, $Xe_{150000}$ wold be assumed to be a spherical plasma that can be treated as a homogeneous sphere, placed in a uniform, time-dependent electric field. The application of the external field polarizes the sphere, and alters the field in and around the sphere (see text).}
\end{figure}

\begin{figure}
\caption{The variation with $Re(\epsilon)$ of the electric field within the dielectric sphere, $E_{in}$, and the polarization pressure, $P_{pol}$. The value of $\epsilon$ was chosen so as to produce a fivefold enhancement of the electric field at resonance. $Im(\epsilon)$ was held constant. The laser intensity was 1$\times$10$^{15}$ W cm$^{-2}$ (see text).}
\end{figure}

\begin{figure}
\caption{Raw output signal from the microchannel plate (MCP) detector near the laser arrival time (t=0). See the text for discussion of the features identified as A, B, and C.}
\end{figure}

\begin{figure}
\caption{Electron energy spectrum from disassembly of $Xe_{25000}$ clusters upon irradiation by 100 fs laser pulses of peak intensity 8$\times$10$^{15}$ W cm$^{-2}$. The laser polarization vector was along the axis of the time-of-flight spectometer.}
\end{figure}

\begin{figure}
\caption{Integrated electron yield from $Xe_{150000}$ as a function of retarding potential on the energy analyzer. The signal for perpendicular polarization has been multiplied by a factor of 2. T$_1$ and T$_2$ are the results of two-temprature fits to the data.}
\end{figure}

\begin{figure}
\caption{MCP signals for electrons for different values of laser pulse duration covering the range from 200 fs to 1200 fs. The laser intensity at 200 fs was 8$\times$10$^{15}$ W cm$^{-2}$ and the laser polarization was along the axis of the TOF spectrometer.}
\end{figure}

\begin{figure}
\caption{Integrated electron yield as a function of laser pulse duration for laser polarization direction that is parallel and perpendicular to the axis of the TOF spectrometer.}
\end{figure}

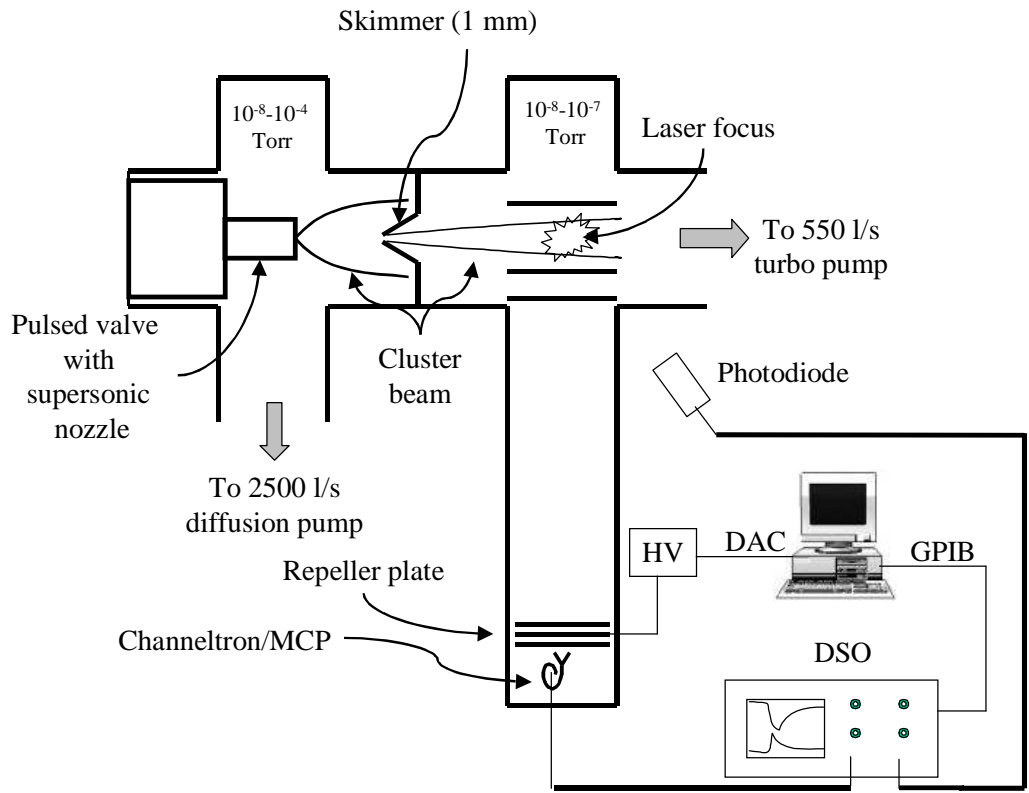
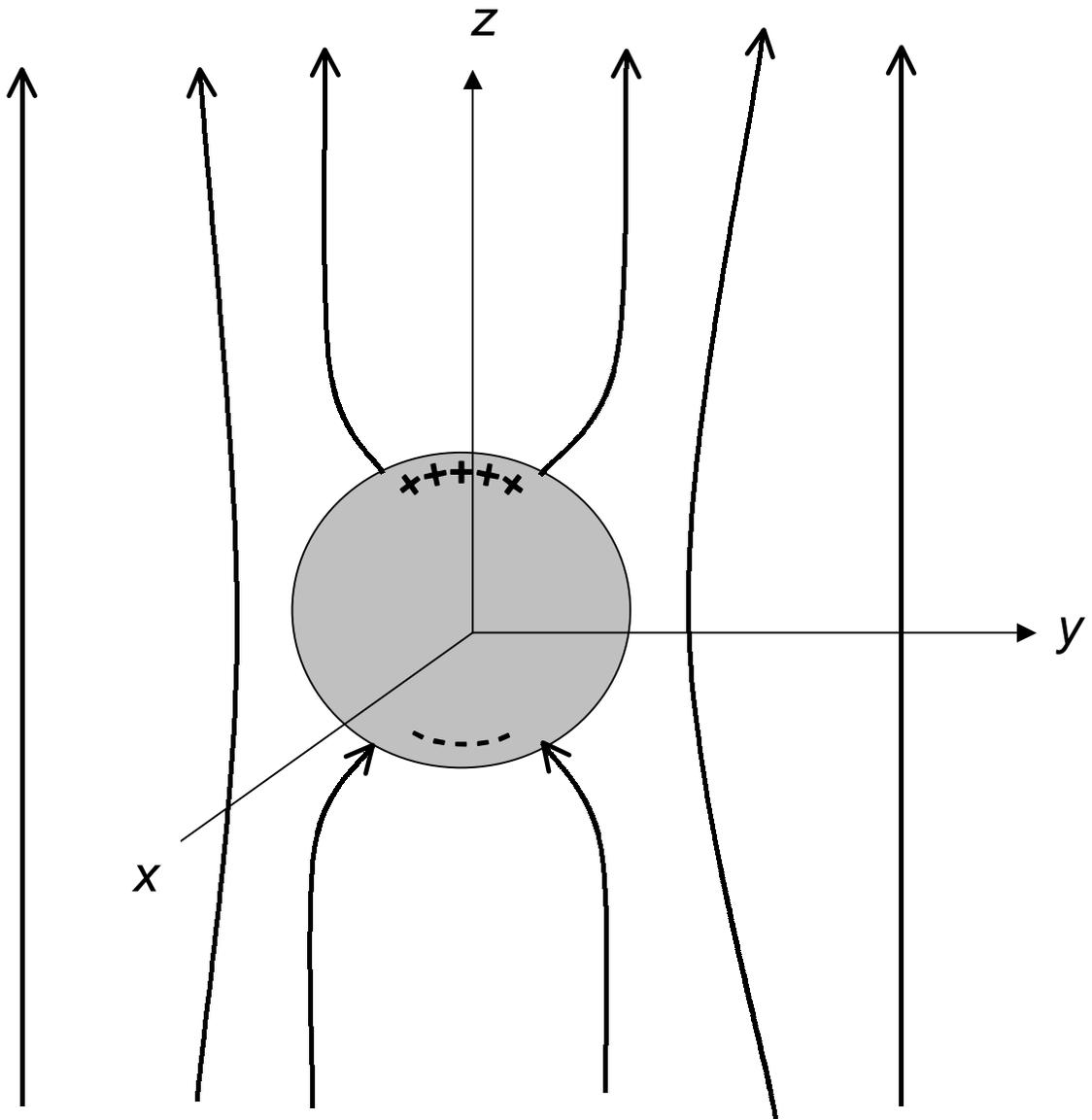
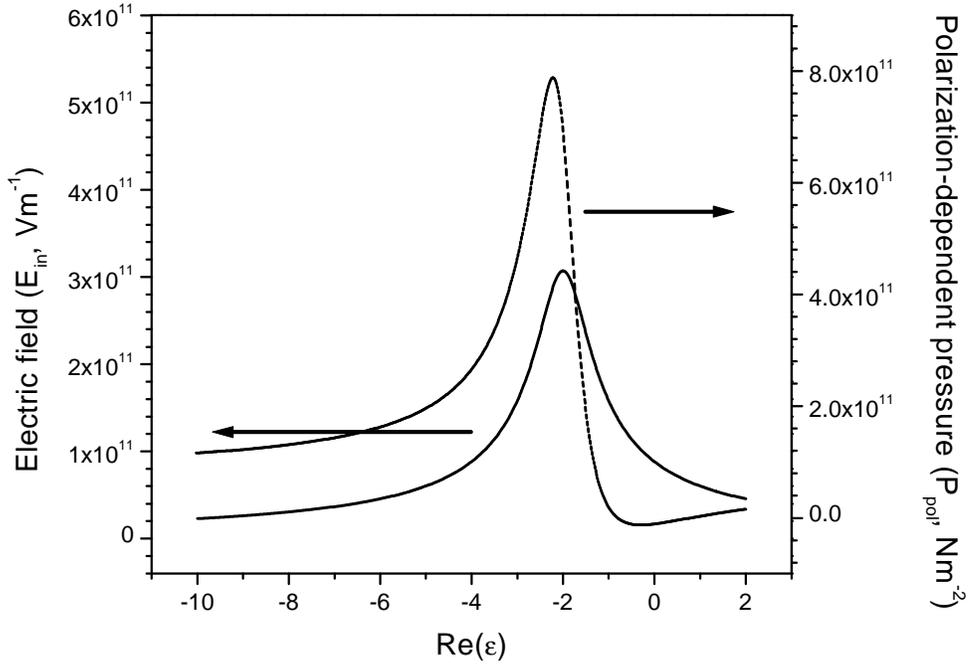
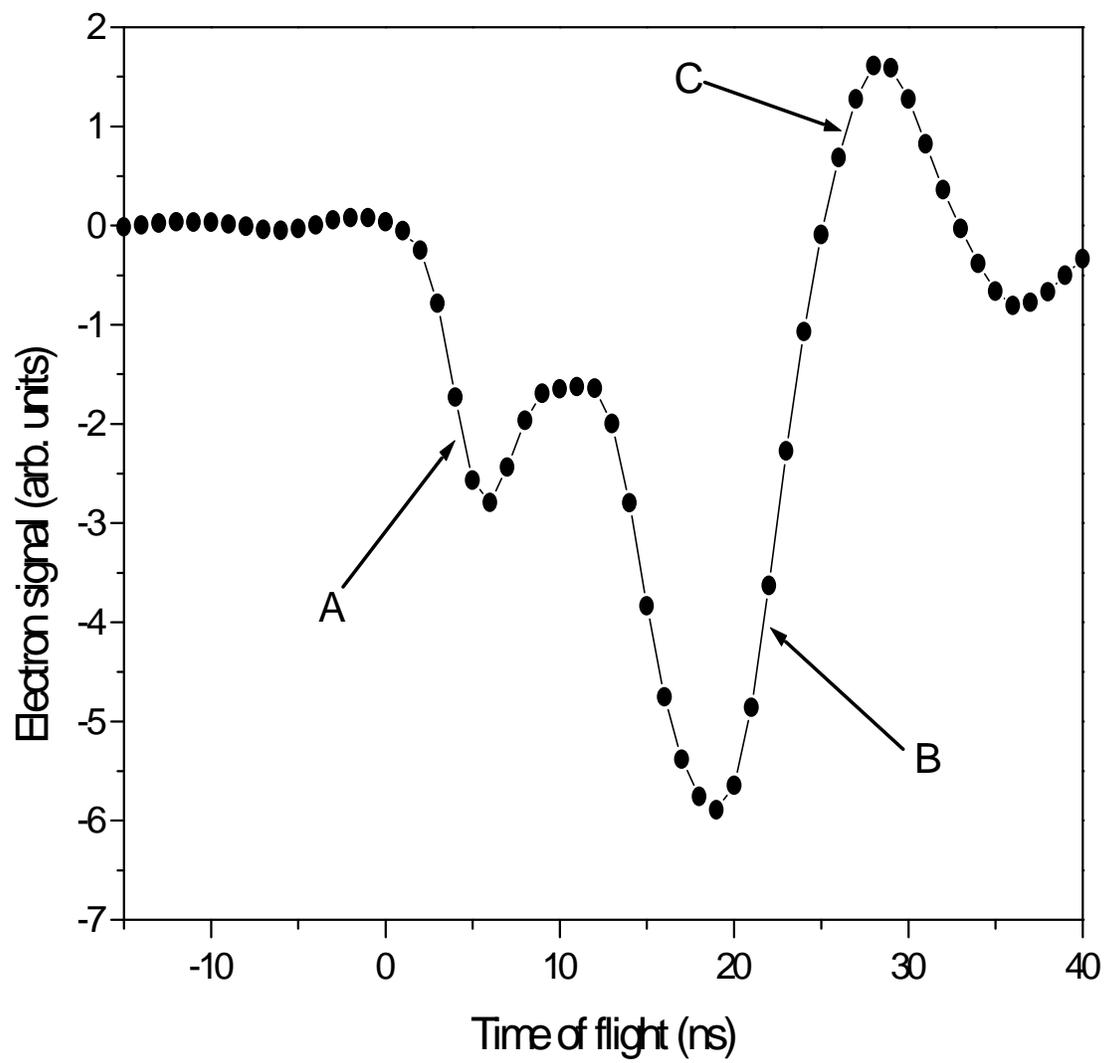
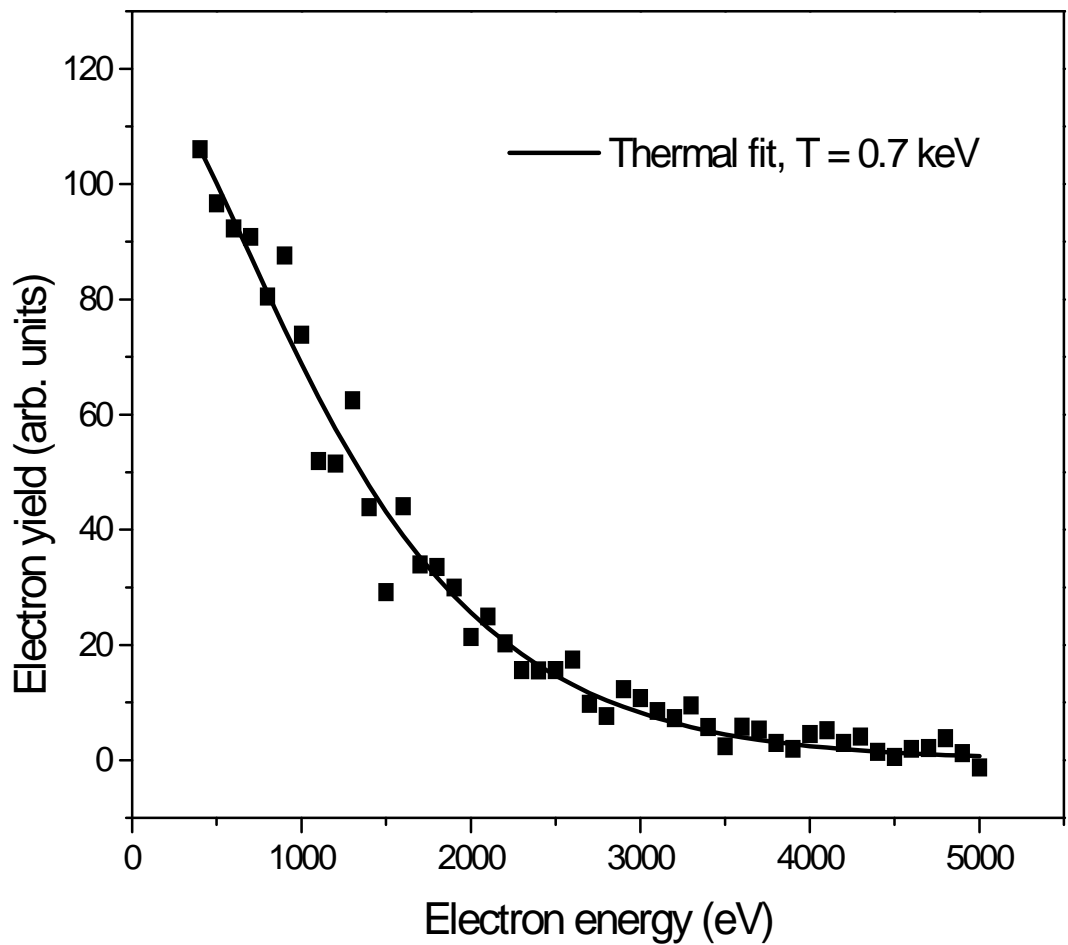
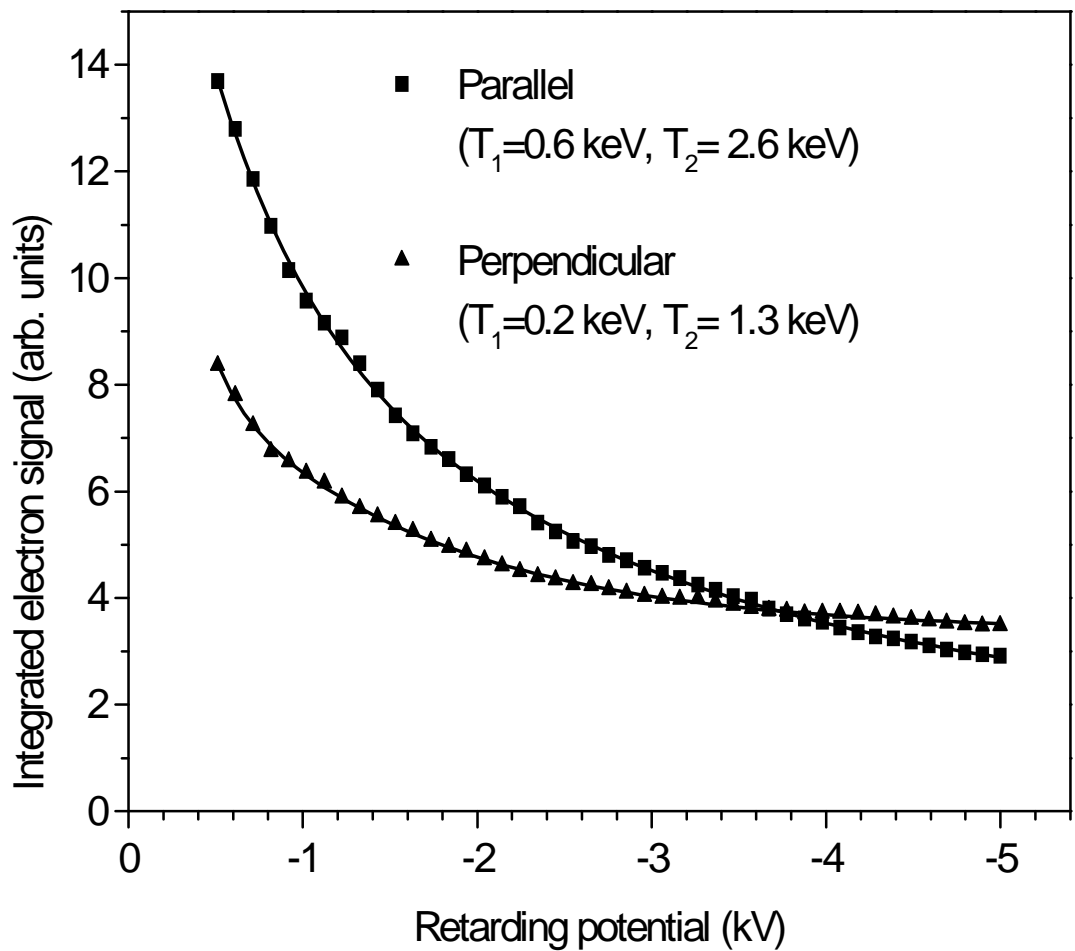
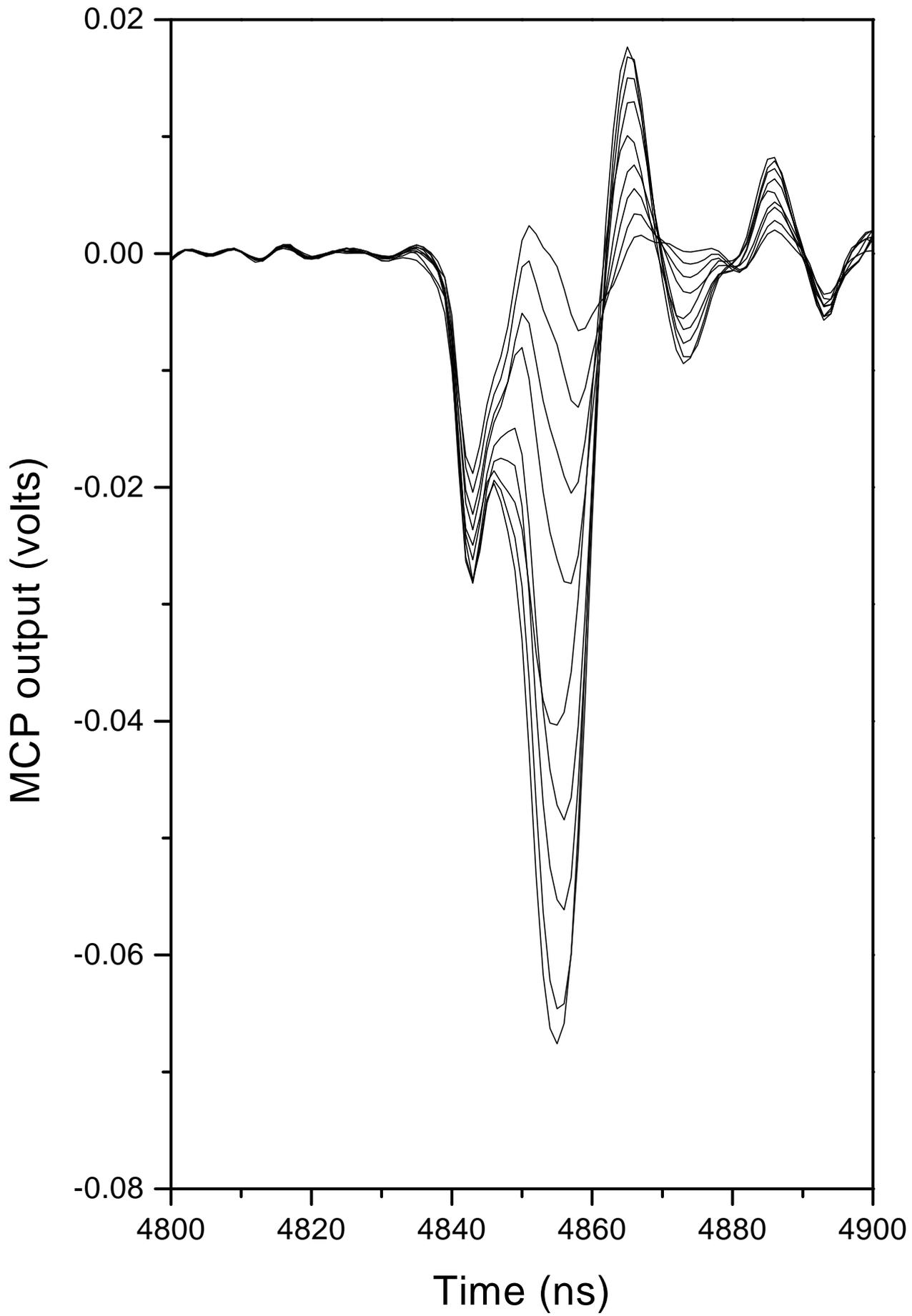
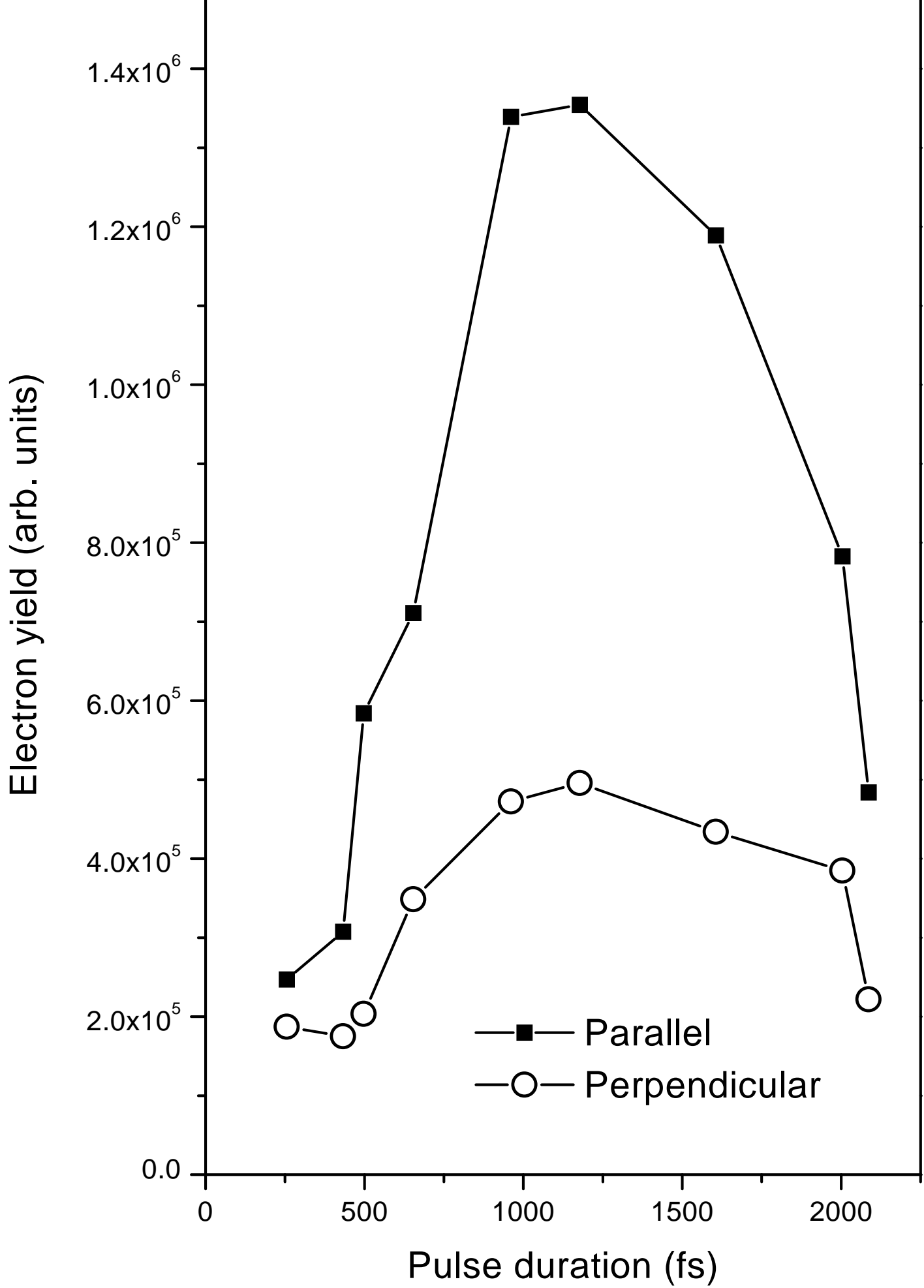
\end{document}